# Network Intrusion Detection Using Wrapper-based Decision Tree for Feature Selection

Mubarak Albarka Umar       Chen Zhanfang       Yan Liu

School of Computer Science and Technology, Changchun University of Science and Technology

7186 Weixing Road, Jilin, China.

12018300037@mails.cust.edu.cn       chenzhanfang@cust.edu.cn       yanl@cust.edu.cn

## ABSTRACT
One of the key challenges of machine learning (ML) based intrusion detection system (IDS) is the expensive computational complexity which is largely due to redundant, incomplete, and irrelevant features contain in the IDS datasets. To overcome such challenge and ensure building an efficient and more accurate IDS models, many researchers utilize preprocessing techniques such as normalization and feature selection in a hybrid modeling approach. In this work, we propose a hybrid IDS modeling approach with an algorithm for feature selection (FS) and another for building an IDS. The FS algorithm is a wrapper-based with a decision tree as the feature evaluator. The propose FS method is used in combination with some selected ML algorithms to build IDS models using the UNSW-NB15 dataset. Some IDS models are built as a baseline in a single modeling approach using the full features of the dataset. We evaluate the effectiveness of our propose method by comparing it with the baseline models and also with state-of-the-art works. Our method achieves the best DR of 97.95% and shown to be quite effective in comparison to state-of-the-art works. We, therefore, recommend its usage especially in IDS modeling with the UNSW-NB15 dataset.

## 1. INTRODUCTION
Protecting the confidentiality, credibility, integrity, and availability of information transferred over the internet and across computers has become a vital and challenging task for many government and private organizations [1]. This prompted the use of an intrusion detection system (IDS) as an added security layer to the already existing less effective first line of defense such as encryption, access control, and firewalls [2], [3]. Machine learning techniques, due to their ability to learn and improve with experience and without much human intervention [4], are nowadays utilized in building such IDS [5]. There was one problem with the initial idea of applying ML in the form of a single classifier in IDS, that is, this approach may not be strong enough to build a good IDS [6]. Thus, to enable building good and more accurate IDS, various researchers proposed hybrid IDS modeling approach in which an algorithm is used as an evaluator in a given feature selection technique in combination with a classifying algorithm to enhance the efficiency and capability of detecting an intrusion by the IDS [7].

IDS datasets are an important aspect in building and validating IDS [8], however, because a realistic IDS dataset typically comes from heterogeneous platforms and can be noisy, redundant, incomplete, and inconsistent [9] with numerous attack types and network traffic attributes, this poses another challenge for ML as they expand the search space of the problem and lead to high computational and time complexity [10], and this generally tends to affects the detection efficiency of an IDS. To overcome such issues, many research recourses to employing some preprocessing techniques such as normalization [11], [12], data filtration [6], discretization [13] and soon. Feature selection is one of these techniques that has been proposed by various researchers for increasing the detection efficiency and accuracy [7] and it has notably proven to be the most effective solution for an IDS [14].

Feature selection is the process of selecting an optimal subset of features that infer the same meaning as the complete feature set and provide the same or an improved IDS performance without transforming the dimension of the features whilst maintain the physical meaning of the dataset [14]. Feature selection aims to select features that are capable of discriminating samples that belong to different classes, and this usually leads to better learning performance (e.g., higher learning accuracy for classification), lower computational cost, and better model interpretability [15]. There are three main feature selection approaches, filter, wrapper, and embedded methods. The filter method extract features from data without using any learning algorithm, it relies on the general characteristics of the data such as distance, consistency, dependency, information, and correlation to evaluate and select feature subsets, it may however eliminate relevant and important features. The wrapper method uses a learning algorithm to determine the most useful and relevant features, and compared to the filter method, it improves model performance but it's computationally more expensive. Due to these limitations in each method, the embedded method was proposed to bridge the gap between the filter and wrapper methods. The embedded model performs feature selection in the learning time. In other words, it achieves model fitting and feature selection simultaneously [15].

In this work, we propose an IDS approach for detecting malicious network traffic with more efficiency and higher accuracy. Wrapper-based decision tree feature selection algorithm is firstly used as a regular means of obtaining an optimal subset of the original features thereby eliminating redundant, repetitive, and unrelated features. Secondly, we selected five among the most used ML algorithms in IDS [5] for building the IDS model, thus, feature selection and any of the selected ML algorithms are combined to build more efficient and accurate IDS with low model building time. The selected algorithms are Artificial Neural Network (ANN), k-Nearest Neighbor (KNN), Support Vector Machine (SVM), Random Forest (RF), and Naïve Bayes (NB). The implementation of these IDS models is conducted using a contemporary and standard UNSW-NB15 dataset [8], [16] introduced by Moustafa and Slay [17]. One-hot encoding and min-max methods are used for encoding and normalization during the implementation process. And beside feature selection's and model's training time (i.e. total execution time), accuracy, detection rate, and false alert rate, the three most used IDS eval metrics [5] are used to evaluate the IDS models. Furthermore, to determine the effectiveness of our method, we performed comparisons with the baseline models and also with the state-of-the-art works. The major contributions of this work are summarized as follows:

I. We propose a hybrid approach towards IDS modeling that combines a feature selection with a classifier and another classifier in implementing the IDS model to increase the accuracy and efficiency of IDS.
II. In the context of feature selection, we propose the use of a wrapper-based approach with BestFirst as the search method and decision tree as the feature evaluator to find the optimal subset of features for IDS modeling.
III. The proposed method is compared with existing methods. Experimental results show that the proposed method surpasses state-of-the-art methods in terms of attack detection rate (ADR) metric. In IDS, the ADR is possibly the most important metrics than the other classification metrics [18].

The rest of the paper is organized as follows. A literature review of the feature selection application in IDS is presented in Section 2. Section 3 introduces the general concept of feature selection, justification of the feature selection choice in this work is also given. Then, the proposed methodology is given in detail in Section 4, while in Section 5 we provide the evaluation results and discussions are made through comparative analysis on performance and model building time basis with the baseline models as well as with state-of-the-art results obtained by researchers in related studies. Finally, the conclusion and future research direction are presented in Section 6.

## 2. LITERATURE REVIEW

Feature selection has been widely used in many domains, such as text categorization [19], genomic analysis [20], [21], intrusion detection [22], [23] and bioinformatics [24]. In this section, we discussed some of the feature selection works within the domain of Intrusion detection.

Sivatha Sindhu et al., [25] proposes a lightweight IDS for multi-class categorization using a wrapper-based genetic algorithm for feature selection and a hybrid of neural network and decision tree (neurotree) for actual classification. They used 16/41 features of NSL-KDD datasets and a min-max method to normalize the selected attributes. WEKA's evaluation measures were used to evaluate the performance of their, and compared to tree-based single classifiers their proposed methods achieved the highest detection rate of 98.38%. Thaseen and Kumar [13] evaluated the classification ability of six distinct tree-based classifiers on the NSL-KDD dataset. They used WEKA's CONS and CFS filters to select 15/41 features of the dataset, however, no normalization was done on the data (possibly because it has no impact on the performance of tree-based algorithms [26]). To evaluate the performance of the models, WEKA's evaluation measures were used and the RandomTree model holds the highest degree of accuracy and reduced false alarm rate. Ghaffari Gotorlar et al., [27] proposed a harmony search-support vector machine (HS-SVM) method for intrusion detection on a KSL-KDD dataset. They used harmony search to select 21/41 best features and the numerical features were normalized using the min-max method whereas the nominal values were converted to numeric. LibSVM library was used for training the SVM model. Detection rate and test time were used to evaluate the model performance, and the results show that the proposed HS-SVM method overcomes the SVM drawback of time-consuming during testing. Khammassi and Krichen [28] proposed the use of three distinct decision tree-based algorithms on a genetic algorithm-logistic regression wrapper selector (GALR-DT) in building IDS models. The three decision tree classifiers used are C4.5, Random Forest, and Naïve Bayes Tree. They applied a wrapper approach based on a genetic algorithm as a search strategy and logistic regression as a learning algorithm to select the best subset of features on KDDcup99 and UNSW-NB15 datasets. 18/41 features were selected in KDDcup99 and 20/42 features were selected in UNSW-NB15 datasets by the GA-LR wrapper. Log-scaling and Min-max of the 0-1 range were applied to normalized the data. Dataset-wise performance of the models was compared using the detection rate, accuracy, and false alert rate. Their results show that UNSW-NB15 provides the lowest FAR with 6.39% and a good classification accuracy compared to KDDcup99 and thus, they conclude that the UNSW-NB15 dataset is more complex than the KDD99 dataset. Setiawan et al., [29] proposed an IDS model using a combination of the feature selection method, normalization, and Support Vector Machine. WEKA's modified rank-based information gain filter was used to select 17/41 NSL-KDD dataset features and the numerical features were log normalized. The model was evaluated using WEKA's evaluation measures and they achieved an overall accuracy of 99.8% Khan et al., [30] proposed a novel two-stage deep learning (TSDL) model, based on a stacked auto-encoder with a soft-max classifier, for efficient network intrusion detection. The model comprises two decision stages and is capable of learning and classifying useful feature representations in a semi-supervised mode. They evaluate the effectiveness of their methods using KDD99 and UNSW-NB15 datasets. DSAE feature selection is used to select 10 features in each dataset, which are normalized using the min-max method. Most used IDS model evaluation metrics were used to assess the performance of their proposed model, achieving high recognition rates of up to 99.996% and 89.134% for the KDD99 and UNSW-NB15 datasets respectively.

The findings from our literature review show that though feature selection has effects on each of the work, certain limitations are found to hinder the vast effects the feature selection process could have had in each of the work. The limitations can be classified into three categories: preprocessing stage (transformation - use of wrong encoding method, superficial feature selection comparison made), modeling stage (use of few algorithms thus, no room for comparison and result validation), and dataset issues (use of outdated dataset). This study aims to overcome these limitations. Since most of the selected ML algorithms in this work consider all features during training simultaneously, one-hot encoding, an approach suited for such ML algorithms [31], is used. Furthermore, min-max, one of the most common normalization method [32], is also used. Five among the widely used ML algorithms in IDS modeling [5] are selected, and a contemporary UNSW-NB15 dataset [33] is selected for this study. The decision tree wrapper-based feature selection approach is used to select the best optimal subsets from the dataset, and a total of ten (10) models are developed and evaluated. In what follows next, the feature selection concept is introduced along with the justification for using the said feature selection method.

## 3. FEATURE SELECTION

In any classification problem, dealing with large datasets may require selecting the most useful and relevant features as many features may contain false correlations, redundant and irrelevant features which can increase computation time, impact the accuracy of the built model. Feature Selection (FS) is one of the important and frequently used data preprocessing techniques for selecting the optimal subset of relevant features from original features for model construction. The optimality of a feature subset is measured by an evaluation criterion [34]. Feature selection reduces the number of features by removing irrelevant, redundant, or noisy data, and brings immediate effects on the subsequently built model [35].

### 3.1 Feature Selection General Procedure

As shown in Figure 1, a typical feature selection process consists of the following four basic steps [34]:

- Subset generation: is a procedure that produces candidate feature subsets for evaluation based on search starting point, direction, and strategy.
- Subset evaluation: this evaluates and compares candidate subset with the previous best one according to a certain evaluation criterion. If the new subset turns out to be better, it replaces the previous best subset. The evaluation criteria can be independent (used in filter methods) or dependent (used in wrapper methods).
- Stopping criterion: this determines when the feature selection process should stop. The process of subset generation and evaluation is repeated until a given stopping criterion is satisfied.
- Result validation: this validates the selected best subset using prior knowledge or different tests by synthetic and/or real-world data sets.

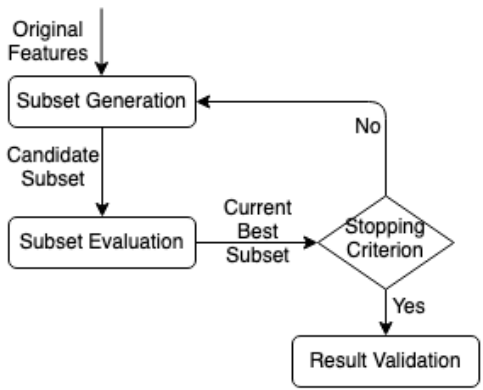

**Figure 1: Feature Selection Procedure**

### 3.2 Feature Selection Methods

Feature selection algorithms designed with different evaluation criteria can be broadly categorized as a filter, wrapper, or hybrid methods [15], [34]. Several researchers have proposed the use of these methods in identifying important features for IDS [35].

- *Filter method*: Separates feature selection from classifier learning so that the bias of a learning algorithm does not interact with the bias of a feature selection algorithm. It relies on the general characteristics of the data to evaluate and select feature subsets.
- *Wrapper method*: Uses the predictive accuracy of a predetermined learning algorithm to determine the quality of selected features. Compared to the filter model, it improves model performance but is prohibitively computationally expensive to run for data with a large number of features.
- *Hybrid method*: Attempts to take advantage of the two methods by exploiting their different evaluation criteria in different search stages and usually achieves comparable accuracy to the wrapper and comparable efficiency to the filter. It first incorporates the statistical criteria, as the filter method does, to select several candidate features subsets with a given cardinality, and then it chooses the subset with the highest classification accuracy as wrapper does. The hybrid methods perform both feature selection and model training simultaneously

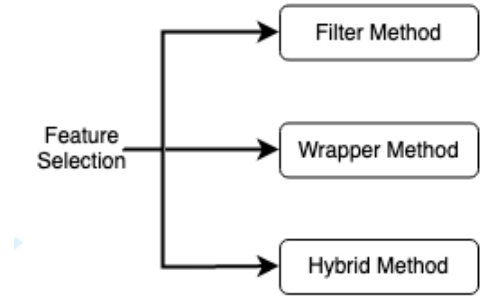

**Figure 2: Feature Selection Methods**

### 3.3 Proposed FS for IDS

Modern intrusion detection datasets inevitably contain plenty of redundant and irrelevant features which can affect the accuracy of IDS model [36], several researchers have tried to handle that using feature selection algorithms to select only the important intrusion features [35]. In this work, we propose the use of a wrapper-based feature selection approach with a decision tree algorithm as the feature evaluator because of the following reasons:

I. Most of the existing IDS datasets contain categorical features [8] and decision tree can handle both categorical and numeric features [26]
II. A decision tree is a low-bias algorithm [37], thus it can select optimal features while avoiding underfitting, which is one of the challenging issues in classification tasks [38]
III. Decision tree can be used to implement a trade-off between the performance of the selected features and the computation time which is required to find a subset [39]. Meaning it can be stopped at any time, providing sub-optimal feature subsets.

The only downside of the proposed method is the expensive computational time, however, giving that IDSs are kind of systems that can be trained offline and deployed online for use [40], [41], this should not be a major point of concern if a satisfactory result is achievable.

## 4. METHODOLOGY

This section described how the experiment is conducted by following the four basic machine learning steps (i.e. data acquisition, data preprocessing, model selection and training, and model evaluation). It also provides the tools used in experimenting.

### 4.1 Experimental Tools

In literature, many tools are used for implementing, evaluating, and comparing various IDS works. WEKA, general-purpose programming languages (such as Java, Python, etc.), and Matlab are the most used tools [5]. In this work, Excel, WEKA, and Python are used for data analysis and exploration, preprocessing, implementing, and validating the IDS models. Jupyter Notebook is used as the execution environment for Python and its libraries.

## 4.2 Dataset Acquisition

In this work, dataset UNSW-NB15, which is among the latest and recommended dataset [8] and is found to be reliable and good for modern-day IDS modeling [16], is used.

### 4.2.1 UNSW-NB15 Dataset

The UNSW-NB15 dataset is a new IDS dataset created at the Australian Center for Cyber Security (ACCS) in 2015. About 2.5 million samples or 100GB of raw data were captured in modern network traffic including normal and attack behaviors and are simulated using the IXIA Perfect Storm tool and a tcpdump tool. 49 features were created using the Argus tool, the Bro-IDS tool, and 12 developed algorithms. The created features can be categorized into five groups: flow features, basic features, content features, time features, and additional generated features. The dataset has nine different modern attack types: Backdoor, DoS, Generic, Reconnaissance, Analysis, Fuzzers, Exploit, Shellcode, and Worms [17]. The UNSW-NB15 is considered as a new benchmark dataset that can be used for IDSs evaluation by the NIDS research community [42] and is recommended by [8]. For easy use and work reproducibility, the UNSW-NB15 comes along with predefined splits of a training set (175,341 samples) and a testing set (82,332 samples) [43], the predefined training and testing sets are used in this work. The publicly available training and testing set both contain only 44 features: 42 attributes and 2 classes. Since our primary focus is binary classification, the broad distribution of total attacks (anomaly) and normal traffic samples of the training and testing sets are used as shown in Table 1.

**Table 1: UNSW-NB15 Distribution Sample**

| Category | Training Set | | Testing Set | |
|---|---|---|---|---|
| | Size | Distribution (%) | Size | Distribution (%) |
| Total Attacks | 119,341 | 68.06 | 45,332 | 55.06 |
| Normal | 56,000 | 31.94 | 37,000 | 44.94 |
| Overall Samples | 175,341 | 100 | 82,332 | 100 |

## 4.3 Data Preprocessing

In this study, two major preprocessing steps are used, namely, data reduction (filtration and feature selection) and data transforming (data normalization and encoding).

### 4.3.1 Data Reduction

#### 4.3.1.1 Data Filtration

Irrelevant data is removed in both the training and testing set. The UNSW-NB15 dataset comes with 42 attributes, 2 class attributes, and an additional id attribute that is removed. Also, since we are only interested in binary classification, the class attribute *attack_cat* indicating the categories of attacks and normal traffic is removed before feature selection.

#### 4.3.1.2 Feature Selection

The training set is used in feature selection to avoiding information leakage and subsequent building of misleading or overfitting models [44], the testing set is solely used for models' performance assessments.

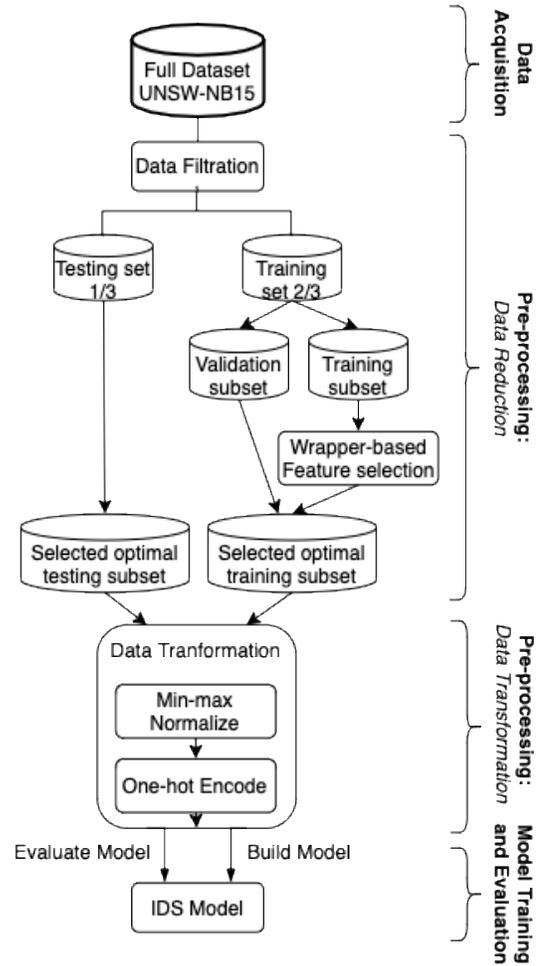

**Figure 3: Wrapper-based FS and Model Training Framework**

We propose the use of a wrapper-based DT approach where *BestFirst* Forward search strategy is used in feature search with 5 consecutive non-improving nodes as the search stopping criteria and accuracy as the evaluation measure. The J48, a java implementation of Quinlan's C4.5 [45], decision tree algorithm [44] provided in WEKA [46] is used as the feature evaluator. Our proposed approach selected 19 optimal features, and Figure 3 depicts the wrapper feature selection and the entire modeling process. After the feature selection operations, WEKA's supervised attribute *Remove* filter is used to collect the subsets of the optimal features. Table 2 shows the selected features.

**Table 2: The Selected Features**

| FS Method | FS No. | Selected Features |
|---|---|---|
| Wrapper-Based DT | 19 | *proto, *service, spkts, sbytes, dbytes, dttl, sloss, dloss, swin, stcpb, trans_depth, response_body_len, ct_srv_src, ct_src_dport_ltm, ct_dst_sport_ltm, ct_dst_src_ltm, ct_flw_http_mthd, ct_src_ltm, ct_srv_dst |
| (*) – indicates nominal features | | |

### 4.3.2 Data Transformation

#### 4.3.2.1 Data Normalization

The UNSW-NB15 dataset consists of two types: numeric and nominal features. To avoid classifier bias towards numeric features with large value ranges, min-max normalization with a range of 0 to 1 is applied on all the numeric features across the datasets using the Equation (1) Below. The normalization process is performed after feature selection in order not to affect the selection process.

$$x_{new} = \frac{x - min(x)}{max(x) - min(x)} \quad (1)$$

#### 4.3.2.2 Data Encoding

All the categorical (nominal) features are one-hot encoded. The full UNSW-NB15 dataset has 39 numeric and 3 nominal features, the nominal features are *proto*, *service*, and *state*. Two nominal features (*proto*, and *service*) are selected by the DT-based wrapper and the Relief filter, whereas the Gain Ration filter selects 18 numeric and only one nominal feature (state). All the 19 features selected by the Information Gain filter are numeric. An example of One-hot encoding of *protocol_type* feature with three sample values is shown in Table 3. Because one-hot encoding increases the dataset dimension, so to avoid losing some nominal features' values encoded during feature selection, the encoding is performed after the feature selection and normalization processes.

**Table 3: One-Hot Encoding Example**

| Protocol_type |   | UDP | TCP | ICMP |
|---|---|---|---|---|
| UDP |   | 1 | 0 | 0 |
| TCP | → | 0 | 1 | 0 |
| ICMP |   | 0 | 0 | 1 |

After encoding the features, the dimensions of the final datasets increased as shown in Table 4. The final encoded features are then used in training the models.

**Table 4: Final Datasets Dimensions**

| Dataset | UNSW-NB15 Dataset features ||
|---|---|---|
|  | Before Encoding | After Encoding |
| Full dataset | 42 | 194 |
| DT Wrapper | 19 | 163 |

## 4.4 Model Selection and Training

Model selection is the selection of a model from many possible trained models using various approaches [47]. In this work, the holdout method, the simplest kind of cross-validation [48], is used. The dataset is separated into two sets, called the training set and testing set. The building of the models constitutes of two stages: training stage and testing stage. During the training stage, the algorithms are trained using the training set, then in the testing stage, the testing set is used to assess the performance of the built IDS models. Figure 3 depicted the entire model training and testing processes. Using the selected algorithms, a total of 10 models are built with the full and the FS datasets. Default Scikit-learn implementation of these algorithms is used in developing the models with SVM's probability set to true as the only altered parameter.

To measure the effectiveness of our method, some evaluation metrics are used to evaluate and compare the models. The model evaluation metrics and the result of the evaluations are provided in the next subsection and the Result and Discussion section of this work respectively.

## 4.5 Model Evaluation Metrics

Most of the IDS works made use of three metrics, namely; classification accuracy, detection rate (DR), and false alarm rate (FAR) [49]. Similarly, in this work, these metrics are adopted in addition to model building time (MBT). The formulas associated with the metrics are shown below.

$$Accuracy(ACC) = \frac{TP + TN}{TP + TN + FP + FN} \quad (2)$$

$$Detection\ Rate(DR) = \frac{TP}{TP + FN} \quad (3)$$

$$False\ Alert\ Rate(FAR) = \frac{FP}{FP + TN} \quad (4)$$

The MBT is the time taken to train and evaluate a model, excluding the FS time. Because the timing depends on factors beyond our control such as CPU task switching, etc., we try to avoid running heavy tasks whilst executing the programs, we also try to prevent the computer from sleeping to ensure minimal interference.

## 5. RESULT AND DISCUSSION

This section presents the platform on which the experiment is performed, the result obtained, and the interpretation of the results. We compare our proposed approach with the baseline models on performance and MBT basis. We also made comparisons with state-of-the-art works.

## 5.1 Experimental Platform

To avoid interference from the experimental platform, all the programs are implemented and executed in the same environment using the same programming language as shown in Table 5.

**Table 5: Experimental Platform**

| Name | Details |
|---|---|
| Computer | Lenovo ThinkPad T450 |
| OS | Windows 7 Ultimate 64-bit |
| CPU | 2.30GHz Intel Core i5 series 5 processor |
| RAM | 8GB (7.70GB usable) |
| Storage Disk | 240GB SSD |
| Execution platform | Jupyter Notebook |
| Experimental Tools | Excel, WEKA, Python |

## 5.2 Comparison with Baseline Models

In this sub-section, comparisons of models built using our proposed method are made with the models built using the full features of the UNSW-NB15 dataset as the baseline. The basis of the comparisons is the performance and the MBT shown in Table 6 and Table 7 respectively.

In ANN models, our method decreases the MBT and performs fairly good, achieving a score that is very close to that of the baseline model on DR and slightly lower on accuracy and FAR. In comparison to the baseline model, our method achieves similar performance with SVM on accuracy and DR metrics. It however has the highest overall MBT, thus it failed to improve the MBT against its counterpart baseline model, and hence, MBT–wise it is the worst IDS model of all. Our method also performed well on the KNN model with scores very close to that of the baseline model across all the metrics and at a considerably lower MBT. The proposed method achieved its best performance on the RF model with an accuracy of 86.41%, which almost the same as that of the baseline model at lower MBT. Finally, with the NB model, our method also achieves similar performance to the baseline model in a lower MBT. In all the models, the NB has the worst DR. Detection rate (DR) is more important than the other metrics in IDS [18], thus performance-wise, it is the worst IDS model.

**Table 6: Models Performance Comparisons**

| Models | Evaluation metrics | UNSW-NB15 | |
|---|---|---|---|
| | | Full Features | DT Wrapper |
| ANN | ACC | 86.00 | 82.08 |
| | DR | 98.62 | 97.94 |
| | FAR | 29.45 | 37.36 |
| SVM | ACC | 81.6 | 79.11 |
| | DR | 99.64 | 99.31 |
| | FAR | 40.51 | 45.64 |
| KNN | ACC | 84.78 | 83.21 |
| | DR | 96.46 | 96.44 |
| | FAR | 29.53 | 33.01 |
| RF | ACC | 86.82 | *86.41* |
| | DR | 98.7 | *97.95* |
| | FAR | 27.74 | *27.73* |
| NB | ACC | 55.61 | 55.61 |
| | DR | 19.39 | 19.38 |
| | FAR | 0.01 | 0.01 |

Overall, the performance of models built using our proposed method, which selected only 19 features, is almost the same as the baseline models which are built using the full features space, this is consistent with the result achieved by [28], furthermore in comparison to the baseline models, our method decreases the MBT of all the models except for SVM model as depicted in Table 7 below. Thus, it can be deduced that our proposed method is promising and can be used to achieve good results in a shorter MBT without using the full features space of the IDS dataset.

**Table 7: Models Building Time Comparisons**

| Dataset Features | Models Building Time (MBT) | | | | |
|---|---|---|---|---|---|
| | ANN | SVM | KNN | RF | NB |
| Full Features | 11.27m | 181.68m | 17.92m | 0.74m | 4.64s |
| DT Wrapper | 4.95m | *259.1m* | 10.94m | 0.63m | 2.86s |

### 5.3 Comparisons with State-of-the-art Works

To examine the effectiveness of our proposed method, we selected the best performing model from among the models implemented using our method for corresponding comparisons with other state-of-the-art IDS work. There are many similar IDS research works, we however limited our comparison to those that also apply feature selection on the UNSW-NB15 dataset. We compare the percentages of accuracy (ACC), attack detection rate (DR), and false alert rate (FAR) whilst also paying attention to feature selection method, number of features, and algorithms used. Table 8 shows the performance comparisons of the works chronologically.

**Table 8: Related Works Comparisons**

| Work 'Year | FS Method | No. | Algorithm | ACC (%) | DR (%) | FAR (%) |
|---|---|---|---|---|---|---|
| [50] '15 | ARM-Based | 11 | LR | 83.0 | 68 | 14.2 |
| [28] '17 | GA-LR | 20 | DT | 81.42 | – | 6.39 |
| [51] '17 | Functional measures | 33 | DL-binomial | 98.99 | 95.84 | **0.56** |
| [30] '19 | DSAE | 10 | DL-Soft-max | 89.13 | – | 0.75 |
| [52] '19 | K-means | 41 | DNN | **99.19** | – | – |
| [53] '20 | NSGAII-ANN | 19 | RF | 94.8 | 94.8 | 6.0 |
| *This Work* | *DT-based* | *19* | *RF* | *86.41* | **97.95** | *27.73* |

From Table 8, it can be seen that our method achieves the best DR of 97.95%, performed better than two methods in ACC, and has the worst FAR. The best ACC and FAR are achieved by [52] and [51] respectively, both of which used deep learning classifiers. Their good results may be influenced by the use of deep learning classifiers which are proving to be good in IDS classification tasks recently [54]–[56]. However, it is important to note that in IDS, not detecting an attack can be costlier than mis-detecting an attack [18], thus DR can be more important than any other evaluation metrics and hence, our method can actually be more effective in detecting an attack than both [52] and [51]. Nonetheless, overall our method can be considered to be quite good and very effective. Its major downside is the expensive computational time required during feature selection and training, however, giving that IDSs are kind of systems that can be trained offline and deployed online for use [40], [41], this should not be a major point of concern since our method is capable of detecting malicious network attack more effectively than the other methods.

### 6. CONCLUSION

It is evident that from the results achieved by our method in comparison to the baseline our method decreases model training and testing or computation time while achieving a performance of almost the same as the baseline models that utilize the full features, thus this depicts the efficiency of our method. Furthermore, in comparison to state-of-the-art works, our method proved to be quite good with it achieving the best attack detection rate. It is therefore sufficient to say that our proposed method is quite effective and we thus recommend its usage in IDS modeling especially when working with the UNSW-NB15 dataset.

One of the major issues observed in all the models developed in this work is the high number of false alert rate, which if any of the models is to be deployed will be rather boring to work with; and besides a high detection rate, a good IDS should have a very low false alert rate, thus more work can be done in the future focusing particularly on reducing these high false alert rate observed.

## 7. ACKNOWLEDGEMENTS

This work was funded by the Scientific and Technological Development Scheme of Jilin Province, People's Republic of China.This work was funded by the Scientific and Technological Development Scheme of Jilin Province, People's Republic of China.